\documentstyle[12pt,aaspp4]{article}

\def\etal{{\em et al.~\/}}

\def\ie{{\em i.e.~\/}}

\lefthead{Mac Low \&  Ferrara}
\righthead{Mass Loss from Dwarf Galaxies}

\begin{document}

\title{Starburst-driven Mass Loss from Dwarf Galaxies:\\
Efficiency and Metal Ejection}

\author{Mordecai-Mark Mac Low$^1$ and Andrea Ferrara$^{2,3}$}
\affil{
$^1$Max-Planck-Institut f\"ur Astronomie, K\"onigstuhl 17,
D-69117 Heidelberg, Germany; Email: mordecai@mpia-hd.mpg.de \\
$^2$Joint Institute for Laboratory Astrophysics, University of
Colorado and National Institute of Standards and Technology, Campus
Box 440, Boulder, CO 80309-0440, USA \\
$^3$Osservatorio Astrofisico di Arcetri, I-50125 Firenze, Italy; 
Email: ferrara@arcetri.astro.it}

\begin{abstract}
We model the effects of repeated supernova explosions from starbursts
in dwarf galaxies on the interstellar medium of these
galaxies, taking into account the gravitational potential of their
dominant dark matter haloes.  We explore supernova rates from one every
30,000 yr to one every 3 million yr, equivalent to steady mechanical
luminosities of $L=0.1-10 \times 10^{38}$~ergs~s$^{-1}$, occurring in
dwarf galaxies with gas masses $M_g=10^6 - 10^9 M_\odot$.  We address
in detail, both analytically and numerically, the following three
questions: 
\begin{enumerate}
\item When do the supernova ejecta blow out of the disk of the galaxy?
\item When blowout occurs, what fraction of the interstellar gas is
blown away, escaping the potential of the galactic halo?
\item What happens to the metals ejected from the massive stars of the
starburst?  Are they retained or blown away?
\end{enumerate}
We give quantitative results for when blowout will or will not occur
in galaxies with $10^6 \leq M_g \leq 10^9 M_\odot$.  Surprisingly, we
find that the mass ejection efficiency is very low for galaxies with
mass $M_g \geq 10^7 M_\odot$.  Only galaxies with $M_g \lesssim 10^6
M_\odot$ have their interstellar gas blown away, and then virtually
independently of $L$.  On the other hand, metals from the supernova
ejecta are accelerated to velocities larger than the escape speed from
the galaxy far more easily than the gas. We find that for $L_{38}=1$, only
about 30\% of the metals are retained by a $10^9 M_\odot$ galaxy, and
virtually none by smaller galaxies. We discuss the
implications of our results for the evolution, metallicity and
observational properties of dwarf galaxies.
\end{abstract}

\keywords{Hydrodynamics -- Shock waves -- Galaxies: Dwarfs 
-- Numerical Methods}

\section{Introduction}

Hierarchical models of structure formation suggest that dwarf galaxies
are the building blocks of larger galaxies, merging at high redshift
to form the distribution of galaxies we see today.  Dwarfs, and even
smaller stellar systems, are also responsible for the origin of
some of the Lyman~$\alpha$ forest absorption features observed in the
spectra of distant quasars (Fransson \& Epstein 1982, Wang 1995,
Ciardi \& Ferrara 1997).  It is therefore crucial to understand both
their formation and evolution.  In turn, it has recently become clear
that supernova-driven winds play a crucial role in the evolution of
such objects since they regulate the mass, metal enrichment, and
energy balance of the interstellar medium (ISM) in these galaxies.
These aspects have been investigated in a number of theoretical papers
(Larson 1974; Dekel \& Silk 1986; Silk \etal 1987; Vader 1986, 1987;
and Ferrara \& Tolstoy 1998, hereafter FT) both in local and
cosmological contexts.

The observational evidence in support of the existence of outflows
from dwarf galaxies has grown rapidly in recent years. Meurer \etal
(1992) found an expanding bubble of ionized gas with $\sim$ kpc size
and velocity of order $100$~km~s$^{-1}$ in the active dwarf galaxy
NGC1705, and Martin (1996) found evidence for superbubbles in I~Zw~18
in which the ionized gas expands with velocities of several tens of
km~$s^{-1}$. Similar results have been obtained by Della Ceca \etal
(1996; see also Heckman \etal 1995) from ASCA observations of
NGC~1569, a star-forming dwarf galaxy in which the authors have been
able to demonstrate the presence of diffuse hot gas ($T\sim 0.7$~keV) very
likely heated by supernova explosions.  Bomans, Chu, \& Hopp (1997)
found X-ray emission from hot gas within a supergiant shell in the
dwarf irregular NGC~4449, which also has a large-scale radio
synchrotron halo (Klein \etal 1996).  In their search for outflows in
dwarf galaxies Marlowe \etal (1995) conclude that the outflow
phenomenona are relatively frequent in centrally star-forming objects;
in addition they point out that these outflows tend to be generally
oriented along the galaxy minor axis.  On a smaller scale, Puche \etal
(1992) in their VLA HI study of dwarf galaxies, have detected
low-density bubbles surrounded by a shell expanding with velocity
$\sim 15$~km~s$^{-1}$. An absorption line study (Bowen \etal 1996) from
low column density gas in the dSph galaxy Leo~1 has given upper limits
on the amount of hydrogen present in the halo of such galaxy ($M \le
10^4 M_\odot$). This low value seems to be in contradiction with the
expectations from simple blowaway models in which the gas content of
the galaxy has been ejected into the halo. However, as the authors
stress, the possibility that a significant fraction of {\em hot} gas
exists along the line of sight cannot be ruled out. (Alternatively,
even if the galactic gas had indeed been ejected, it could have been
stripped either by ram pressure against the intergalactic medium (IGM) or by
a galactic encounter.)

In many cases these observed gas velocities clearly exceed the escape
velocity from the potential well of the parent galaxy, and therefore
this material will be injected into the IGM and lost to the
galaxy. However, it is difficult to determine what fraction of the
ambient medium is indeed brought to such high speeds.  Mass can be
ejected from a galactic disk through the ``blowout'' of a superbubble
driven by multiple supernova remnants, carving a hole through the
galactic disk (Mac Low \& McCray 1988).  Gas accelerated above the
escape velocity is then ``blown away'', and escapes the potential of
the galaxy entirely (De Young \& Heckman 1994). Determining the
conditions under which blowout occurs, and under which most of the ISM
can be blown away are necessary steps towards the understanding of the
evolution of such objects, and also of the possible relationships
between early and late type dwarfs.

In particular, for blowout there is yet no clear answer to the
question of what fraction of the gas swept up in the shell is
accelerated above the escape velocity. The reason for this is that
most previous studies in which blowout has been simulated
numerically (Tomisaka \& Ikeuchi 1986, 1988; Mac Low \etal 1989;
Tenorio-Tagle, Rozyczka, \& Bodenheimer 1990; Silich 1992; Mineshige,
Shibata, \& Shapiro 1993, Suchkov \etal 1994) have not been followed
for long enough evolutionary times. Moreover, these models have been
almost totally concerned with superbubbles and starbursts in massive
galaxies; thus it is not straightforward to extrapolate their results
to dwarfs, which have rather different properties, both in terms of
their ISM and, perhaps more important, of their dark matter content.

De Young \& Gallagher (1990) did model a dwarf galaxy undergoing a
starburst, and concluded that about $1/6$ of the mass of the swept-up
shell can escape the galaxy.  However, their study neglected the
presence of dark matter in the galaxy, and assumed a thin gaseous disk
with a scale height of $150$~pc, as well as suffering from
insufficient numerical resolution, leaving their conclusions
questionable.
Ideally, one would also like to find how the fraction of mass that can
escape depends on the metallicity of the gas; the cooling provided by
metals can be strongly inhibited in the relatively unpolluted gas of
dwarfs, whose metallicity is typically more than one order of
magnitude below solar (Skillman, Kennicutt, \& Hodge 1989).

A related important point concerns the fate of metals ejected by
supernovae exploding in the burst. The perceptive idea introduced by
Vader (1986) that outflows can be ``metal enhanced'', although
appealing, needs to be quantitatively examined.  De Young \& Gallagher
(1990) did try to investigate the fate of metals in their model.  They
concluded that most of the metal-rich supernova ejecta are indeed
ejected, but that as much as 1/3 of the ejecta end
up within the swept-up shell surrounding the hot bubble.  However,
this appears to have been due to unphysical diffusion of their tracer
particles from the hot interior gas of the bubble into the surrounding
cold shell, due to the low numerical resolution of their model.  If
this result were valid, then the concept of metal-enhanced winds would
be much less viable, since the metal-to-gas ratio in the wind could
not achieve values much higher than in the ambient ISM of the galaxy.
Given the apparent problems with the computation, however, it seems
worthwhile to reexamine this issue.

In summary, the present study concentrates on dwarf galaxies, and
attempts to answer the following questions: 
\begin{enumerate}
\item What are the conditions for either blowout or blowaway to occur (see our
definitions in \S~3)?
\item What fraction of gas escapes the galaxy when they do?
\item What is the fate of metals ejected by the central supernovae?
\end{enumerate}
We investigate these problems taking into account the full
structure of dwarf galaxies, by making physically meaningful models
for the structural properties of dwarf galaxies in order to determine
the characteristics of the ambient medium in which the multiple
supernova explosions take place. We combine analytic calculations with
numerical hydrodynamic simulations using ZEUS-3D\footnote{Developed by
the Laboratory for Computational Astrophysics at the National Center
for Supercomputer Applications, and available for community use by
registration at the email address lca@ncsa.uiuc.edu}, a code using the
algorithms described by Stone \& Norman (1992).

The plan of the paper is as follows. In \S~2 we describe our model for
the structure of dwarf galaxies, based on
recent data on their dark matter and gas content. In \S~3 we describe
how we perform our numerical computations. Before presenting the 
results of these experiments in \S~5, we derive in \S~4 simple 
analytical conditions for the occurrence of blowout and blowaway as a function 
of the mass of the galaxy and the mechanical luminosity of the central
supernovae. \S~6 contains discussion and a brief summary of the paper.  

\section{Structure of Dwarf Galaxies}

For our purposes, we model a dwarf galaxy as a system made of a
visible disk with combined
gaseous and stellar mass $M_g$, and a dark matter
halo of mass $M_h$.  We assume that the gas has a density distribution
given by $\rho_g(\varpi,z,\phi)=\rho_0 f(\varpi,z)$, where
$\rho_0=\rho(0,0)$, $\varpi$ is the galactocentric radius, $z$ is the
vertical coordinate, and $f(\varpi,z)$ is a function determined, as
described below, by imposing hydrostatic equilibrium on the gas in the
total gravitational potential of the dwarf galaxy,
$\Phi_t(\varpi,z)=\Phi_h(\varpi,z)+\Phi_g(\varpi,z),$
which consists of halo and disk components.

The behavior of the dark-to-visible mass ratio $\phi=M_{h}/M_g$
in galaxies has been explored in great detail in a key study by Persic \etal 
(1996). These authors find that $\phi$ is a function of the galactic
mass; using their relations one can easily derive the dependence of
this ratio on the visible mass of the galaxy:
\begin{equation}
\label{phi}
\phi \simeq  34.7 M_{g,7}^{-0.29}.
\end{equation}
where $M_{g,7}=M_g/10^7~M_\odot$. From this equation it is clear that
the gravitational potential of dwarf galaxies with $M_g \lesssim 10^9
M_\odot$ is dominated by the dark matter halo; therefore we will
neglect the potential due to visible mass $\Phi_g$ in all the
following calculations involving the total potential $\Phi_t$.  The
general trend of increasing dark matter fraction with decreasing
galaxy mass is indeed consistent with other kinematical studies of
dwarf spheroidal galaxies (Mateo 1997) for which velocity dispersions
of a significant number of stars have been derived. However, we
caution that equation~(\ref{phi}) was derived by Persic \etal (1996)
from a sample of larger galaxies with $M_h \gtrsim 10^{10} M_\odot$ or
$M_g \gtrsim 10^{9} M_\odot$.  Lacking more suitable data,
we are extrapolating those results to lower masses. To make this
assumption clear throughout the paper we will keep $\phi$ indicated in
a general form as long as possible.  

The density profiles of the dark matter haloes remains uncertain.
Cold dark matter cosmological models predict that halo
density profiles are essentially self-similar with a weak mass
dependence (Lacey \& Cole 1994; Navarro \etal 1997); specifically the
dark matter density should decrease as $\rho_h^{CDM}(r) \propto
r^{-1}$ at small radii, where $r^2=\varpi^2 + z^2$.  The detailed form
of $\rho_h^{CDM}(r)$ is found to be (Navarro \etal 1997)	 
\begin{equation}
\label{rhocdm}
\rho_h^{CDM}(r) = 1500 \rho_c
\frac{r_{\rm 200}^3}{r (5r + r_{\rm 200})^2},
\end{equation}
where $\rho_c$ is the central density and $r_{200}$ is the
characteristic radius within which the mean dark matter density is 200
times the present critical density, $\rho_{crit}=3 H_0^2\Omega/8\pi
G=1.88\times 10^{-29} h^{-2} $~g~cm$^{-3}$, where $H_0= 100
h$~km~s$^{-1}$~Mpc$^{-1}$ is the Hubble constant; throughout the paper
we will use $\Omega=1$.  Equation~(\ref{rhocdm}) holds down to $r_{min} \sim
10^{-2} r_{ 200}$~kpc, where the resolution of the simulations becomes
inadequate.

Recent observational work, however, appears to disagree with this
prediction.  Salucci \& Persic (1997) have demonstrated convincingly
that dark matter halos have a core, \ie a central region of almost
constant density, whose size increases with galaxy luminosity both in
absolute units and as a fraction of the optical radius.  The structure
of very small galaxies with mass $\sim 10^7 M_\odot$ is even less
well understood.  In view of the uncertain state of the art,
we calculate the halo gravitational potential by assuming that the density
distribution of the halo can be approximated by a modified isothermal
sphere (Binney \& Tremaine 1987), which is general enough to be
appropriate for an idealized situation such as the one presented here, and
does reproduce the observed central core.  It follows that
\begin{equation}
\label{rhoh}
\rho_h(r)= {\rho_c \over 1+ (r/r_0)^2}.
\end{equation}
The  halo mass as a function of radius is then
\begin{equation}
\label{mhr}
M_h(r)=\int_0^r dr' 4\pi r'^2\rho_h(r') = 4\pi\rho_c r_0^3\left(x -
\arctan x\right),
\end{equation}
where $x=r/r_0$. Thus, if $x\gg 1$, which we will show later on,
\begin{equation}
\label{mh}
M_h = M_h(r_h) \simeq 4\pi\rho_c r_0^2 r_h, 
\end{equation}
where $r_h$ is an appropriately defined halo radius. Following a
common {\em Ansatz} we take 
\begin{equation}
\label{rh}
r_h\equiv r_{200}= \left(\frac{3\rho_c}{200 \rho_{crit}}\right)^{1/2} r_0,
\end{equation}
as obtained using the definition of $r_{200}$ given above and
equation~(\ref{mh}).  To proceed further we need a relation between
the scale radius, $r_0$, the central dark matter density $\rho_c$, and
the mass of the halo $M_h$.  Burkert (1995) has shown that the total
dark matter inside $r_0$, given by $M_0=M_h(r_0) = \pi(4-\pi)\rho_c
r_0^3$ (which implies $M_0 \simeq 0.21 (r_0/r_h) M_h$), is related to $r_0$ and
$\rho_c$ through the relations
\begin{eqnarray}
\label{grelsa}
M_0 & = 7.2\times 10^7 \left({r_0\over {\rm
kpc}}\right)^{7/3}~~M_\odot, \\
\rho_c & = 2.7\times 10^7 \left({r_0\over {\rm
kpc}}\right)^{-2/3}~~M_\odot~{\rm kpc}^{-3}. \label{grelsb}
\end{eqnarray}
Substituting for $M_0$ into equations~(\ref{grelsa}) and~(\ref{grelsb}) we get 
\begin{eqnarray}
\label{r0}
r_0 & = & 4\times 10^{-7} \left(\frac{M_h}{M_\odot}\right)^{3/4} 
                          \left(\frac{r_h}{1 \mbox{ kpc}}\right)^{-3/4}
                          \mbox{ kpc}, \\
\label{rhoc}
\rho_c & = & 5\times 10^{11} \left(\frac{M_h}{M_\odot}\right)^{-1/2}
                             \left(\frac{r_h}{1 \mbox{ kpc}}\right)^{1/2} 
                             M_\odot \mbox{ kpc}^{-3}.
\end{eqnarray}
Substituting the value of $r_h$ from equation~(\ref{rh}) into these
expressions we obtain the dependence of the halo radius on $M_h$, 
\begin{equation}
\label{rh1}
r_h= 0.016 \left({M_h\over M_\odot}\right)^{1/3} h^{-2/3} {\rm kpc},
\end{equation}
and can substitute back into equations~(\ref{r0}) and~(\ref{rhoc}), to
find
\begin{eqnarray}
\label{r01}
r_0 & = &5.3\times 10^{-5} \left({M_h\over M_\odot}\right)^{1/2}
h^{1/2}  \mbox{ kpc}, \\
\label{rhoc1}
\rho_c &=& 2\times 10^{10} \left({M_h\over M_\odot}\right)^{-1/3}
h^{-1/3}  M_\odot \mbox{ kpc}^{-3}.
\end{eqnarray}
With these assumptions the gravitational potential of the halo is 
\begin{equation}
\label{grav}
\Phi_h(r) = 4\pi G \rho_c r_0^2 \left({1\over 2} \log \left(1+ x^2\right) + 
{\arctan x\over x}\right).
\end{equation}
Thus the galaxy has a circular velocity 
\begin{equation}
\label{vc}
v_c^2(r) = r {\partial \Phi_h\over\partial r}= {4\pi G\rho_c r_0^2\over x}
(x-\arctan x);
\end{equation}
The rotation curve increases rapidly in the inner parts of the galaxy,
already being practically flat at $x=1.5$.  The asymptotic value of
the rotation velocity at $x \gg 1$ can then be found by using
equations~(\ref{r01}) and~(\ref{rhoc1}) to be
\begin{equation}
v_c = (11.85 \mbox{ km s}^{-1}) (M_{g,7} h \phi)^{1/3}.
\end{equation}
We tabulate the circular velocities for galaxies of various masses in
Table~\ref{pars}, taking $h = 0.65$ and using equation~(\ref{phi}) for
the mass ratio $\phi$.

We then compute the gas density distribution by solving the steady state 
momentum equation, 
\begin{equation}
\label{hydroeq}
({\bf v \cdot \nabla}){\bf v} = -{1\over \rho}{\bf \nabla} P - {\bf \nabla} \Phi_h,
\end{equation}
where $P=\rho c_{s,eff}^2$, $\rho$ is the gas density,
$c_{s,eff}^2=c_{s}^2 + \sigma^2$ is an effective sound speed, $c_s$ is the 
gas sound speed, and $\sigma$ its turbulent velocity.  To compute
$r_0$, we again assume $h = 0.65$.
In the horizontal direction, the gas is supported against the gravitational 
pull by both thermal pressure and rotation.
In analogy with the dark matter component,
we suppose that the gas extends out to a disk cut-off radius, $\varpi_*$. 
The latter quantity
is obtained from a fit to a sample of dwarfs by FT. They  find that
the radius-HI mass relation is well approximated by the law
\begin{equation}
\label{rc}
\varpi_* \simeq  \varpi_0 M_{g,7}^\alpha=3  M_{g,7}^{0.338}~~~ {\rm
kpc},
\end{equation}
defining the constants $\varpi_0$ and $\alpha$.
Cutting the disk at an arbitrary radius does introduce a region out of
hydrostatic equilibrium; as a result the disk tends to expand at its
own effective sound speed during the simulations.  As this is rather a
low velocity, it only results in a slight rounding of the sharply
cut off outer edges of the disks.

We will use equation~(\ref{rc}) thoughout the paper, although one should be
aware of the presence of some degree of uncertainty.  For the sake of
clarity, it might be useful to give the explicit expressions for the
relations among the three characteristic radii so far introduced in
this Section. These are:
\begin{eqnarray}
\label{rrels}
{r_h \over \varpi_*} &\simeq& 1.23 (\phi h^{-2})^{1/3}\\
{\varpi_* \over r_0} &\simeq& 15.6 M_{g,7}^{-1/6}(\phi h)^{-1/2}\\
{r_h \over r_0}      &\simeq& 19.2 M_{g,7}^{-1/6}(\phi h^{7})^{-1/6}.
\end{eqnarray}
The last equation justifies the approximation $x\gg 1$ used in eq. \ref{mh}.

Outside $\varpi_*$ we postulate the existence of an IGM
whose present density, $\rho_{IGM}= \Omega_b\rho_{crit}$ has been
estimated by assuming $\Omega_b=0.02$ and $h = 0.6$,
giving $\rho_{IGM} \sim  1.4 \times 10^{-31}$~g~cm$^{-3}$.
When $\rho_g(\varpi,z)$ has decreased to $\rho_{IGM}$, we set it constant
for the remainder of the grid.  As this gas is not in hydrostatic
equilibrium, it tends to accrete onto the galaxy. This efffect is
negligible: if the entire grid were filled with gas of this
density, it would have a mass of only $\sim 10^5 M_\odot$, and if it
accreted at a constant rate over a typical 200 Myr run, it would give
an accretion rate of only $5 \times 10^{-4} M_\odot$~yr$^{-1}$.

With the above assumptions, the solution of equation~(\ref{hydroeq})
provides a reasonably self-consistent initial distribution for the gas
in which supernova explosions are simulated.  The derived distribution
is uniform to 1\% in the equatorial plane of the galaxy.  A cut along
the vertical axis through the galactic center shows that a
Gaussian distribution matches the actual distribution very well.

However, in order to provide in the simplest possible manner a
tractable analytic estimate for the order of magnitude of the scale
height $H$, $n_0=\rho_0/\mu m_h$, and $N_{H}$, the gas column density,
we make the assumption that all the mass is concentrated in a thin
disk, producing a constant gravitational acceleration $g= 2\pi G
\Sigma_t$, where $\Sigma_t=(M_d+M_h)/2\pi \varpi_*^2$ is the total
matter surface density, using the prescription for the ratio of dark
to visible matter $\phi$ given by equation~(\ref{phi}).  Such a
potential would generate an exponential gas distribution with parameters
given by the following formulae:
\begin{eqnarray}
\label{nhn}
H&=& {c_{s,eff}^2\over 2\pi G \Sigma_t}={c_{s,eff}^2 \varpi^2\over G
M_g(1+\phi)}\simeq 
2 \left({\varpi_0^2 \over \phi}\right) M_{g,7}^{2\alpha-1}
c_{10}^{2}~~~{\rm kpc},                                    \\
\label{n0}
n_0&= &{M_g\over 2\pi H \varpi^2 \mu m_h}=   
2\times 10^{-2} \left({\phi\over \varpi_0^4}\right)
M_{g,7}^{2(1-2\alpha)} c_{10}^{-2}~~~{\rm cm}^{-3},        \\
N_H &=& n_0 H=2\times 10^{20} M_{g,7}^{(1-2\alpha)}
\varpi_0^{-2}~~~{\rm cm}^{-2}, 
\end{eqnarray}
where, $c_{10}=c_{s,eff} /{\rm 10~km~s}$, $\varpi_0$ is defined by
equation~(\ref{rc}) in kpc, and $\mu$ is the mean molecular weight. We
tabulate in Table~\ref{pars} these values as a function of $M_g$ for
the given values of $c_{s,eff}$.  We use only $n_0$ from these values
in the simulations, but they are useful for analytic estimates.
We note that in this simple approximation, $N_H$ is independent of both
$\phi$ and $c_{s,eff}$ and, given that $1-2\alpha \simeq 1/3$, only
weakly dependent on the visible mass of the galaxy. 

\section{Analytical Considerations and Estimates}
\label{analytic}

Before we present the results of the numerical simulations,
it is instructive to derive simple analytical relations that
approximately give the final fate of the shocked gas and of
the galaxy itself. 

We distinguish between two potential results of a central starburst.
First, in a ``blowout'', the central supernova explosions blow a hole
through the galactic gas distribution, parallel to the steepest
density gradient (usually along the rotation axis), accelerating some
fraction of the gas and releasing the energy of subsequent explosions
without major effects on the remaining gas.  Second, in a
``blowaway'', all, or nearly all, of the ambient ISM is accelerated above
the escape velocity and is lost to the galactic
potential well.  In the following we derive in brief the
conditions under which the two processes can occur, as a
function of the mechanical luminosity of the starburst, and of the
mass of the galaxy. A more extended discussion can be found in FT.

\subsection{Blowout} 

The blowout condition can be derived by requiring that
the blowout velocity $v_b$ exceeds the escape velocity of the
galaxy $v_e$.  The blowout velocity can be defined as the velocity
at a height $z=3H$ above the galactic midplane, where $H$ is the
exponential scale height. Since the velocity of a shock produced
by an explosion in a stratified medium decreases down to a minimum
occurring at $z \approx 3H$ before being reaccelerated in case of blowout, 
the above definition of $v_b$ corresponds to the minimum of such a
curve.
The blowout condition follows from an
analysis of the Kompaneets (1957) solution for an explosion in a
stratified medium. The explicit expression for $v_b$ has been obtained
by FT:
\begin{equation}
\label{vb}  
v_b = {e^{3/2}\over 3^{2/3}} \left({125\over 154 \pi}\right)^{-1/6}
\left({L\over \rho_0}\right)^{1/3} H^{-2/3}.
\end{equation}
We recall that the gas mass for a vertically exponential,
horizontally constant density distribution is 
\begin{equation}
\label{mg}
M_g=2\pi \varpi_*^2 \mu m_h n_0 \int_0^\infty dz e^{-z/H}=2\pi \varpi_*^2
\mu m_h n_0 H.
\end{equation}
Using equations~(\ref{nhn}), it follows that
\begin{equation}
\label{vb1}
v_b = 2.7 (4\pi G L)^{1/3} (1+\phi)^{1/3} c_{s,eff}^{-2/3}.
\end{equation}
In general the effective sound speed $c_{s,eff}$ has a contribution from a
turbulent velocity, but we make no attempt to model its value, merely
assuming that it is constant.

Eq. \ref{vb1} can now be cast in the convenient form
\begin{equation}
\label{vb2} 
v_b= 92  L_{38}^{1/3}(1+\phi)^{1/3} c_{10}^{-2/3}~~~{\rm km~s}^{-1}.
\end{equation}
We note that the value of the blowout velocity does not depend 
on the radius $\varpi_*$, but only on the the mechanical luminosity of
the explosion and the dark-to-visible mass ratio.

Now we can compute the escape velocity $v_e$ at the disk radius for
our comparison.  It is
\begin{equation}
\label{ve}  
v_e^2(\varpi_*) = 2 \vert \Phi_h(\varpi_*) \vert \sim 8\pi G \rho_c
r_0^2 \left[{1\over 2}  
\log (1+ x_*^2) + {\arctan x_*\over x_*}\right] ~~~x_*=\varpi_*/r0.
\end{equation}
Taking the appropriate value for $x_*$, as derived from
eq. \ref{rrels}, the terms in parenthesis give a factor $\simeq 1.65$
for all galactic gas masses, since it is $x_* \gg 1$. It follows that
\begin{equation}
\label{ve1}  
v_e(\varpi_*) \sim (13.2\pi G \rho_c)^{1/2} r_0  \sim 20
M_{g,7}^{1/3}(\phi h)^{1/3}~~~{\rm km~s}^{-1},
\end{equation}
where we've used equation~(\ref{phi}) for the relation between the
dark-to-visible mass ratio and the galactic mass, and $h$ is the
scaled Hubble constant.  Again, we remark that $v_e$ is basically
independent of $\varpi_*$ as long as $x_*\gg 1$, which implies an
almost flat $v_e(\varpi)$.  Finally, we can compare $v_e$ to the blowout
velocity $v_b$ to find the condition for the blowout to occur:
\begin{equation}
\label{L38}
L_{38} > 1.2 \times 10^{-2} M_{g,7} c_{10}^2 h.
\end{equation}

\subsection{Blowaway} 

We now derive the necessary condition for a starburst to blow away
(completely unbind) the ambient gas of the galaxy. 
We have seen that blowout takes place when 
the shell velocity at $z=3H$ exceeds the escape velocity.
Following the blowout the pressure inside the cavity drops
suddenly due to the inward propagation of a rarefaction wave.
The lateral walls of the shell, moving along the major axis of
the galaxy, will continue to be accelerated by the interior pressure
until they are reached by the rarefaction wave at $\varpi=\varpi_c$,
corresponding to a time $t_c$ elapsed from the blowout.
After that moment, the shell enters the momentum-conserving phase,
since the driving pressure has been dissipated by the blowout.
The requirement for the blowaway to take place is then that the
momentum of the shell (of mass $M_c$ at $\varpi_c$) is larger
than the momentum necessary to accelerate the gas outside
$\varpi_c$ (of mass $M_o$)at a velocity larger than the escape
velocity:
\begin{equation}
\label{momentum}
M_c v_{\varpi_c} \ge M_o v_e,
\end{equation}
With some simple assumptions and algebra outlined in FT, one can 
write the condition for the blowaway to occur:
\begin{equation}
\label{cond}
{v_b\over v_e}\ge (\epsilon - a)^2 a^{-2} e^{3/2}, 
\end{equation}
where the radius of the shell in the plane when the pressure is
released is $\varpi_b = aH \sim (2/3)H$, and $\epsilon=\varpi_*/H$ is
the ratio of the major to the minor axis of the galaxy. 
Using the expressions given in \S~2 for $\varpi_*$ and $H$, we find
\begin{equation}
\label{eps}
\epsilon= 0.43 {\phi\over \varpi_0}M_{g,7}^{1-\alpha} c_{10}^{-2}.
\end{equation}
Flatter galaxies (larger $\epsilon$ values) preferentially undergo
blowout, whereas rounder ones are more likely to be blown away.  The
condition for the blow away to occur, obtained expanding
eq. \ref{cond} assuming that $\epsilon \gg a$, is then
\begin{equation}
\label{L38a}
L_{38} > 8  \times 10^{-2} M_{g,7}^{7-6\alpha} \left({\phi\over \varpi_0}\right)^{6}
c_{10}^{-10} h.
\end{equation}
We point out the strong dependence of blowaway on the amount of turbulence
present in the galactic ISM: as the value of $c_{10}$ is increased, less powerful
explosions can lead to complete disruption of the galaxy. This is because virialized
objects with higher effective sound speed tend to be rounder, and hence more 
fragile with respect to blowaway. The same type of situation might occur if
for some reason the distribution of the gas is stretched and becomes less
concentrated, as, for example, after the first episode of
star formation, or following a galactic encounter. Thus after a first stage of
the evolution in which blowout dominates, blow away might occur if the
gas distribution can be puffed up enough.

\subsection{Final Fate of the Galaxy}

From the previous results we can predict the final fate of the galactic
gas after a starburst episode. In Figure 1 we show the different regions
of the $M_g-L_{38}$ plane in which blowout or blowaway occurs. 
In this Figure we have used equation~(\ref{rc}) for $\varpi_*$,
equation~(\ref{phi}) for $\phi$, and equation~(\ref{eps}) for
$\epsilon$. Galaxies with gas masses below $M_{g,7}=0.04-0.1$ 
undergo blowaway almost independently of the values of the mechanical luminosity 
of the starburst in the range $L_{38}=0.1-10$. The blowout has a somewhat
steeper dependence on $M_{g,7}$: for the luminosities  $L_{38}=0.1-10$,
blowout can take place up to masses of $\sim 10^{8}-10^{10} M_\odot$, respectively.
This, as we will see in the next Sections, is in excellent agreement with the
results of our numerical simulations. 

\section{Initial Conditions of the Simulations}

For our numerical models, we use ZEUS-3D, a second-order,
Eulerian, astrophysical, gas dynamics code (Stone \& Norman 1992) using
Van Leer (1977) monotonic advection.  

We have implemented equilibrium radiative cooling using the cooling
curve given by MacDonald \& Bailey (1981), which is a Raymond \& Smith
(1977) cooling curve extended to lower temperatures.  We use an
implicit energy equation, implemented with a Newton-Raphson root
finder, supplemented by a binary search algorithm for occasional zones
where the tabular nature of the cooling curve prevents the
Newton-Raphson algorithm from converging.  We also include an
empirical heating function tuned to balance the cooling in the
background atmosphere, but linearly proportional to density, so that
it is overwhelmed by cooling in compressed gas (Mac Low \etal 1989).
This is to prevent the background atmosphere from spontaneously
cooling, and may be thought of physically as a crude model for the
stellar energy input into the background atmosphere.

We use a tracer field advected with exactly the same algorithm as the
density to follow the metal-rich gas ejected by the
central energy source.  The absence of this field then serves to
specify the regions in which cooling should be effective.  This is
particularly useful in this problem for preventing spurious cooling of
the bubble interior due to density numerically diffused off the thin,
dense shell of swept up gas surrounding the bubble.  In order to
maintain a sharp interface at the edge of the hot bubble, we use the
method suggested by Yabe \& Xiao (1993), which consists of advecting a
function $f(c)$ of the tracer field rather than the tracer field $c$
itself.  We follow Yabe \& Xiao in using the function
\begin{equation}
f(c) =  \tan[0.99 \pi(c-0.5)], ~~~~~~~~~c =  0.5 + [\arctan f(c)] / (0.99 \pi).
\end{equation}
With this definition, $c$ will show a sharp transition at the
interface if it has initial values of 0 and 1, even if $f(c)$ becomes
quite smooth due to numerical diffusion.  The factor of 0.99 is used
to avoid infinite values for $c =0$ or $c=1$.  This formulation of the
tracer field strongly reduces numerical diffusion of ejected material
into the cold shell as compared to the tracer particles used by De
Young \& Gallagher (1990).

The gas distribution of the galaxy is set up in hydrostatic
equilibrium with the dark halo potential given in
equation~(\ref{grav}, where $\rho_c$ is given by
equation~(\ref{grelsa}).  To find $r_0$, we use equation~(\ref{r01}),
taking $h = 0.65$, and proceeding as described following those
equations. 

The central energy source is set up as a constant luminosity wind
driven by a thermal energy source.  We note that Strickland \& Stevens
(1998) have shown that a single central energy source will more
efficiently transfer energy to the ambient gas than multiple smaller
clusters with the same total luminosity, as we describe in more detail
in \S~6.  A source region of radius five zones and volume $V$ is set
up initially in pressure equilibrium with the background atmosphere,
but at a temperature a factor $\eta = 1000$ larger than that of the
background atmosphere, and density correspondingly lower by a factor
of $1/\eta$.  Every timestep we then added energy to the source region
at a rate $\dot{E} = L/V$, and mass at a rate $\dot{M} = \rho_0
\dot{E} / \eta e_0$, where $e_0$ is the midplane energy density of the
background atmosphere, maintaining roughly constant specific energy in
the source region, and generating a supersonically expanding wind.
The amount of mass in this wind is significantly greater than that
expected from supernova ejecta alone, as we are trying to also account
for mass that would have evaporated off the shell walls had we
included thermal conduction in our model.  This is only a very rough
approximation; as we discuss below, the resulting temperatures are, in
fact, somewhat cooler than they would be realistically, which gives us
a good lower bound on the metal ejection rates that we compute.  We
chose this energy input phase to last 50~Myr in our models, the
typical lifetime of the least massive star able to become a supernova
(e.g., McCray \& Kafatos 1987), effectively assuming simultaneous
formation of all the stars in the starburst.

We assume azimuthal symmetry and so use in each of our models a
two-dimensional grid with the center of the galaxy at the origin, with
reflecting boundary conditions along the symmetry axis and along the
galaxy mid-plane, and outflow boundary conditions on the other two
axes.  


\section{Numerical Results}

We have performed a parameter study at a resolution of 25 pc per zone,
varying the gas mass $M_g$ of the dwarf galaxies from $10^6$
M$_{\odot}$ to $10^9$ M$_{\odot}$ (see Table~\ref{pars} for equivalent
rotational velocities and halo masses), and varying the mechanical
luminosity of winds and supernovae from the central starbusts from $L
= 10^{37}$ ergs~s$^{-1}$ to $10^{39}$ erg~s$^{-1}$.  These mechanical
luminosities are equivalent to a supernova with energy of $10^{51}$
erg exploding every 3~million yr to every 30,000 yr, respectively.  In
this Section we report the results of these models, describing their
overall development, computing the efficiency of mass and metal
ejection, and making some rough approximations to their observable
emission.  The results in this section on the occurrence of blowout
and blowaway agree very well with the analytic results described in
\S~\ref{analytic} and summarized in Figure~1.

\subsection{Gas Distribution}

We begin by showing in Figure~\ref{den} a time history
of the density distribution of all the models in the parameter study.
The first part of the Figure shows the state of
the gas at a time of 50 Myr, at the end of the energy input phase.
In the higher luminosity models, the internal termination shock can
be clearly seen, surrounded by a hot, pressurized region of shocked
wind and supernova ejecta.  This is in turn surrounded by a colder,
radiatively cooled shell of swept-up ambient gas from the galaxy and
surrounding IGM.  (Another minor effect that can be observed is that
the ambient gas at the edge of each galaxy, which is not pressure
confined, also begins to expand into the surrounding medium at its
sound speed of 10 km/s, producing the rounded ends of the disks.)  The
larger bubbles have begun to accelerate, and show strong
Rayleigh-Taylor instabilities.  

Our assumption of azimuthal symmetry has the consequences described in
Mac Low {\em et al.} (1989) for the Rayleigh-Taylor instabilities.  In
three dimensions, these instabilities produce a characteristic spike
and bubble morphology.  In azimuthal symmetry, only the central spike
can grow, while the other modes are forced into rings.  As a result,
the central spike is more pronounced than it would be in three
dimensions, and the other spikes are less numerous.  However, Mac Low
{\em et al.}  (1989) showed that changing to an assumption of slab
symmetry ({\em e.\ g.} using Cartesian coordinates) did not prevent
the appearance of a central spike.  Although the instabilities do not
fully develop because of these effects, they do function to
effectively allow hot gas to accelerate beyond the colder, denser,
swept-up shells.

After energy input ceases, the holes in the planes of the galaxies
recollapse under the influence of gravity and the pressure of the disk
gas, except in the extreme cases of low mass and high mechanical
luminosity, where the disk gas escapes the potential of the dark
matter halo, and is swept completely off the grid at late times.  At
the same time, the hot bubbles rising above the disks expand out into
the surrounding IGM, leaving most of the swept-up shell material
behind.  After 200 Myr, at the last time shown, the galactic disks
have almost completely recovered unless blow-away occurred.
Meanwhile, low density gas originating in the central winds and
supernovae spreads over regions of tens of kpc.  In Figure~\ref{col}
we show the distribution of metal-enriched wind and supernova ejecta
material after 200 Myr for the more massive models.  In all cases in
which blow-out occurs, this material escapes from the disks of the
galaxies; much of it is travelling at high enough velocity to escape
from the halo potential as well.

This result should be fairly independent of the details of how the hot
gas is traced, so long as hot gas is not allowed to unphysically cool
or diffuse into the cold shell.  Cooling can happen by numerical
diffusion transferring too much mass into the hot interior from the
cold shell, as occurred in the models of Tomisaka \& Ikeuchi (1986),
for example, while diffusion can happen if tracer particles diverge
away from the gas they should be following, as occurred in De Young \&
Gallagher (1990).  Physically, the hot gas will not radiatively cool
within a dynamical time, and its sound speed is significantly higher
than the escape velocity from any of our galaxies, so most of it
should indeed escape as is seen in our models.

\subsection{Mass and Metal Ejection Efficiencies}

We can directly compute the efficiency of mass ejection and retention
of metal-enriched gas from our models.  The question we ask is how
much of each is travelling at speeds higher than the local escape velocity
due to the potential of the dark matter halo.  In principle, the
efficiency of mass ejection is 
\begin{equation} \label{effeq}
\xi = (M_{\rm esc} + M_{\rm lost}) / (M_{\rm i} + M_{\rm sn}), 
\end{equation}
where $M_{\rm esc}$ is the mass on the grid moving at higher than the
escape velocity, $M_{\rm lost}$ is the mass lost off the edge of the
grid, $M_{\rm i}$ is the mass initially on the grid at $t = 0$, and
$M_{\rm sn}$ is the mass injected at the center of the grid by winds
and supernovae during the first 50 Myr.  To get a good value for $\xi$
we must measure it after most of the gas has been accelerated to its
final velocity but before it has left the grid.  We chose a value of
100 Myr as a compromise.  Although the shells from the lowest mass
galaxies have already left the grid, they were moving at above the
escape velocity as they went off, so it is reasonable to assume that
all mass lost from the grid escaped. We also need to make special
provision for the tendency of the relatively large disks in the most
massive models to expand sideways off the grid at the sound speed.  As
no high-speed gas has left the grid in these models, we do this by
neglecting $M_{\rm lost}$ for these models.  In Table~\ref{mass} we
show the ejection efficiency $\xi$ for each of our models.  Only in
our most extreme models, with masses of $10^6$~M$_{\odot}$, is most of
the mass ejected.  In more massive objects, less than 7\% of the mass
is ejected, usually far less.

The tracer field allows us to separately determine the fate of the
metal-enriched gas ejected by the massive stellar winds and SNe of the
starburst.  We compute the efficiency of ejection of this material
$\xi_Z$ by applying equation~(\ref{effeq}) only to the gas carrying
the tracer field.  In Table~\ref{metal} we give $\xi_Z$ for each
model.  In the more massive galaxies and at lower supernova rates, a
significant fraction of the metal enriched gas is retained in the
gravitational well of the dark halo, and will eventually fall back on
to the galaxy, while at lower masses and higher luminosities,
virtually all of the metals escape the grasp of the halo and travel
freely into the surrounding intergalactic space.

When estimating the metal ejection efficiency $\xi_Z$, we have assumed
that the metals ejected by massive stars and supernovae are well mixed
within the hot interior of the bubble, and are therefore ejected
uniformly with this gas.  Two questions can be raised here.  First,
will metals ejected by later supernovae mix with the hot gas of the
already existing bubble?  A simple order of magnitude calculation
(Tenorio-Tagle 1996) shows that the metal diffusion time in the hot
bubble gas (mostly consisting of evaporated shell material) is $t_d
\sim 10^7 n_{-2} R_{100}^2 T_6^{5/2}$~yr, where $n_{-2}=n/10^{-2}{\rm
cm}^{-3}$ is the electron density, $R_{100}=R/100$~pc is the radius of
the bubble, and $T_6=T/10^6$~K is the temperature of the hot gas. This
time is comparable with, but shorter than the evolutionary timescale
of the superbubbles studied in the previous Section.

Second, how will the mixing proceed if SN ejecta are strongly clumped?
This might well be the case if Rayleigh-Taylor instabilities act
during expansion of the shock through the outer layers of the star
(Fryxell \etal 1991).  Such high-density, high-metallicity clumps
could travel almost unimpeded inside the hot, rarefied interior of the
bubble, but would be rapidly thermalized when they hit the external
shell.  In this case the metals would be mixed into the shell, rather
than into the escaping hot gas. The conditions under which the clumps
can reach the shell, rather than being destroyed in the superbubble
interior, have been derived by Franco \etal (1993). These authors
found that for reasonable density contrasts in the clumps, only the
clumps produced by the first 5--10 SNs in the starburst can reach the
shell and pollute it with their metal content. The values of $L_{38}=$
0.1--10 and the duration of the energy input phase considered in this
paper of 50 Myr imply $\sim$ 15--150 SNs.  Thus, ejecta clumping can
play a role in the low-luminosity starburts, where we expect that a
significant fraction of the metals contained in ejecta clumps ends up
not in the hot gas, but rather mixed with the cool shell gas. For
these objects our $\xi_Z$ might represent an overestimate; this error
should become fairly small for higher luminosity starbursts. It
has to be kept in mind, though, that the clumping factor of the ejecta
is still far from well-established.
 
We have checked that our results are reasonably well converged numerically
by rerunning one of our models with $M_g = 10^9 M_{\odot}$ and $L =
10^{38}$ erg s$^{-1}$ at a resolution of 40 zones per scale height,
twice that of our standard models.  We find that, although details of
the shell structure differ slightly, the overall morphology remains
almost constant, and the computed efficiencies shift by no more than
10-20\%, little enough for our conclusions to remain valid.

\subsection{Observables}
 
Finally, we discuss some predictions of observable
properties that we can draw from our models.  As our assumptions about
the microphysics of the gas are quite crude, we are only in a position
to make qualitative predictions; however we believe these can
nevertheless be useful.  Furthermore, we find ourselves in basic
agreement with the more careful work performed by Suchkov et
al.\ (1994) for massive starbursts, adding credibility to our
approximations. 

We begin by computing the temperatures on our grid from the specific
energies computed by ZEUS.  To do this, we need to assume a
composition for the gas, so we take it to be singly ionized with one
atom of He for every 10 protons.  Figure~\ref{temp} shows the
resulting temperatures (for the more massive models).  The hot gas on
the grid appears somewhat cooler than physically reasonable.  This is
most likely due to our replacement of the effects of conduction from
the cold shell with the injection of an arbitrary amount of gas at the
center of the superbubble as discussed in \S~3.

These temperatures can then be used to make rough approximations to
the emissivity in X-rays and in optical lines such as H$\alpha$.  For
optical emissivity, we take 
\begin{equation}
I_{opt} \propto \rho^2 T^{-1/2},
\end{equation}
while for X-ray emission, we take
\begin{eqnarray}
 I_x &  \propto & \left\{  \begin{array}{ll} 
                        \rho^2 \sqrt{T} & \mbox{if $T > T_{min}$} \\
                             0          & \mbox{otherwise}
                           \end{array} \right. 
\end{eqnarray}
We chose $T_{min}$ to have an unrealistically low value of $10^5$~K in
order to more clearly delineate the hot gas emissivity.  This choice
may be justified for the reasons outlined above, but it should be
remembered that these are only qualitative images of the distribution
of emissivity.

In Figure~\ref{opt} we show the optical emissivity, while in
Figure~\ref{xray} we show the X-ray emissivity.  Note that these are
two-dimensional slices across the galaxy, so they would have to be rotated
around the axis and tilted to the line of sight before being directly
compared to observations.  However, they do clearly indicate the
regions expected to be bright in line emission and in X-ray emission.
Probably the most important thing to note in this Figure is the clear
separation between the relatively dense filaments and the bulk of the
hot gas.  Deep imaging of optical lines often shows filamentary or
bubble-like structures in these dwarf galaxies (e.g.\ Marlowe et
al. 1995; Della Ceca \etal 1996); as can be seen from Figure~\ref{opt}
these usually do not trace the outer edges of bubbles but often
represent instead fragments of cold shell left behind during the
blowout of hot gas to much greater distances.

\section{Conclusions}

We have explored the effects of stellar winds and SN explosions from
starbursts with mechanical luminosities ranging from $10^{37}$ to
$10^{39}$~ergs~s$^{-1}$ on the ISM of dwarf galaxies with mass $10^6$
to $10^9 M_\odot$.  Specifically, we have addressed in detail the
following three questions: What are the conditions for blowout and
blowaway to occur, as defined in \S~3?  What fraction of
gas escapes the galaxy in a blowout episode?  What is the fate
of metals ejected by the massive stars during their lives and in their
terminal SN explosions?

We have studied the problem by means of both analytical and numerical
techniques, checking --- where possible --- the agreement between the
two types of results. We have found that dwarf galaxies in the above
mass range undergo blowout for moderate-to-high luminosity values
($L_{38}\sim 1-10$) whereas blowout is inhibited for galaxies with
$M\gtrsim 10^8 M_\odot$ if $L_{38}\simeq 0.1$, as shown in
Figure~\ref{den} and described in Tables~\ref{mass}
and~\ref{metal}. However, the fraction, $\xi$, of the gas mass of the
galaxy lost in such an outflow is surprisingly low in all studied
cases: we find that more than a few percent of the mass is lost only
in objects with masses as low as $10^6 M_{\odot}$.  In larger objects,
the fractions drop to as low as $10^{-6}$ (Table~\ref{mass}).  In the
lowest mass objects, the blowaway occurs virtually independently of
$L$. Thus it appears that explosive mass loss is a process involving a
threshold determined by the geometry of the gas distribution as well
as the depth of the potential well.  The coupling between the energy
input and the disk gas becomes better for the smaller galaxies as they
are more spherical and they need less energy input to begin with for
blowaway to occur.  It is worth stressing that our results are
dependant on the total mass of the galaxy, or the gravitational
potential in which the gas is embedded.  Although we have tried to
model the dark halo carefully, using the best available observational
and theoretical results, the precise value of the dark-to-visible mass
ratio and the details of dark matter distribution remain quite
uncertain.  Our general results should be rather solid, but the
specific values we quote for efficiency and the mass threshold for
blowaway can only be taken as representative.

On the other hand, metal-enriched material from stellar winds and SN
ejecta is far more easily accelerated than the ambient gas to
velocities above the escape speed.  We find that for $L_{38}=1$ less
than 3\% of the metal-enriched material is ejected from a galaxy with $M_g
= 10^9 M_\odot$ of visible mass, but this fraction increases to 60\% for
$M_g = 10^8 M_\odot$, and to unity for $M_g=10^7 M_\odot$ (see
Table~\ref{metal}).  That is, the smaller galaxies inject virtually
all metals produced by the massive stars of the starburst into the
intracluster medium or IGM.  Stronger starbursts with higher
mechanical luminosities eject metal-enriched material even more
easily; only for lower starburst luminosities do galaxies retain a
large fraction of the heavy elements produced by massive stars.

Our assumption that all energy injection occurs in the central hundred
parsecs of the galaxy can questioned, as real starburst galaxies have
multiple clusters scattered across the disk, which one might think
could be more effective at blowing away the ISM than a single, central
energy source.  However, Strickland \& Stevens (1998) have shown with
a simple but elegant argument that our assumption is actually the one
leading to the most effective blowaway for a given total luminosity.
The gist of their argument can be seen by considering a superbubble in
a homogenous medium (Weaver et al. 1977), whose radius $R \propto
L^{1/5} t^{3/5}$, but whose kinetic energy $E \propto L t$.  The
coupling of a superbubble to the background ISM can be characterized
by the energy per unit volume supplied by the superbubble to the ISM
$E/V \propto L^{2/5} t^{-4/5}$, which drops with decreasing
luminosity.  Thus, several small bubbles will be less effective at
transferring energy to the ambient gas than one large one, and so
multiple clusters will be less effective at blowing away the ISM than
our assumed single central cluster.

These results have several implications for the evolution of dwarfs.
First, outflows from dwarf galaxies should be strongly metal enriched.
Tables~\ref{mass} and \ref{metal} show that the typical metal fraction
of the outflow greatly exceeds that of a homogeneously mixed galactic
disk.  This enrichment occurs because the metals produced by the
massive stars in the starburst remain within the hot, shocked cavity
gas, which has a high sound speed and so can easily escape the
galaxy. On the other hand, it is much more difficult to accelerate the
denser, cooler shell of swept-up ambient gas, so that gas remains
bound.  De Young \& Gallagher (1990) found that about 2/3 of the SN
ejecta in the chimney escape from the galaxy for a case in which
$L_{38}=0.3$ and $M_g \sim 10^9 M_\odot$.  For similar parameters,
Table \ref{metal} shows that the blowout is prevented, due to the
presence of the dark matter halo and a realistic hydrostatic
distribution of the gas resulting in a larger scale height (about
three times larger than theirs). This suggests that dark matter is key
to the evolution of dwarf galaxies and any future models must include
it properly.
 
Our simulations show that after blowout has occurred, a substantial
fraction of the hot, metal-rich gas escapes from the galaxy. Metals
are then dispersed into the IGM (subject to the caveats discussed in
the previous section). As a consequence, it appears that dwarfs could be
the major pollutors of the IGM, and certainly have major effects on
the environment in which they live.  We note the result by Renzini
(1997), who finds that most of the iron in clusters appears to reside in the
intracluster gas rather than inside galaxies, a clear sign
that outflow phenomena are at work in that environment.  It is out of
the scope of this paper to review the vast literature concerning these
issues; for a detailed discussion see FT. We only mention that several
authors have already argued that dwarf galaxies might be the most
important objects for the IGM enrichment: our calculations provide a
solid basis for these previous works and should allow the development of
more detailed scenarios.

Starbursts in dwarf galaxies may be traced and studied in detail by
X-ray and optical emission-line studies. The results presented here
show that huge gaseous halos, with sizes of dozens of kpc are
produced, with regions of high x-ray emissivity close to the galactic
disk.  Relatively cool, dense filaments also occur near the galaxy,
well within the external shock, due to shell fragmentation.  High
spatial resolution spectra of the filaments may be useful to
investigate the radiation field in the halo of dwarfs and the escape
fraction of ionizing photons from massive stars in the disk, since
this cool gas should be predominantly photoionized and hence show up
in optical emission lines.  Since these filaments are surrounded by
the hot gas, the X-ray emission may actually be strongest in regions
close to the filaments, where evaporation takes place.  Thus, the bulk
of the observed X-ray emission may come from these conductive
interfaces, in which the gas is far out of ionization equilibrium, and
hence emitting strongly. (These effects are not included in
Figure~\ref{xray}). The filling factor and the nonequilibrium state of
such regions have to be taken into account properly in order to avoid
overestimates of the mass and energy involved in the X-ray emission.

The temperature of the hot gas ($T \sim 0.7$~keV) detected by Della
Ceca \etal (1996) in their observations of NGC~1569 is in qualitative
agreement with the ones shown in Figure~\ref{temp}. We have not tried
to apply our models to a specific object; an interpretation of the
soft X-ray emission from the dwarf galaxy NGC~5253 in terms of a
superbubble model has been attempted by Martin \& Kennicutt
(1995). They concluded that a simple superbubble model with
$L_{38}=1$, exploding in a smooth ambient medium would underproduce
the observed X-ray luminosity, $L_x \sim 6.5\times
10^{38}$~ergs~s$^{-1}$, by a factor $\sim 16$. As an alternative, they
suggest that a clumpy ambient medium generating extra mass of hot gas
via cloud evaporation, may help reconcile the above discrepancy.  We
emphasize that the required hot material could come from the
evaporation of the shell fragments seen in our simulations, without
the need for a multi-phase medium, whose existence in dwarf galaxies
has still to be proven.  Due to their large number, dwarf starbursting
galaxies may significantly contribute to the X-ray background (Persic
\etal 1989), although a precise estimate crucially depends on their
X-ray luminosity function.

The halos created by the starburst energy injection might also be
detectable via absorption line studies towards background objects such
as QSOs. Of course this type of experiment is limited by the low
column densities in the large, rarefied halo, as well as the
availability of suitably placed background objects, which could be a
rare occurence for the small dwarf galaxies.  A particularly suitable
case has been investigated by Bowen \etal (1996).  The dSph Leo~1
has three QSOs in the line of sight though the halo; however, the authors
were only able to put upper limits on the presence of ionized species
like CIV, SiV and MgII, but not to exclude the presence of a hotter
diffuse component.  Similar studies would be particularly important
for dIrr and BCD galaxies which are much more active and gas rich and,
in general, for normal galaxies.

Finally, we recall that the blowout and blowaway episodes discussed
here might be able to carry magnetic field lines along
into the halo.  It is reasonable to expect that, if a sufficiently
strong magnetic field does exist in dwarf galaxies, extended
radio-halos should be found around such objects. Radio continuum
observations of dwarf galaxies are still scarce; probably the most
complete survey has been presented in the pioneering work by Klein
\etal 1984, who concluded that, on average, blue compact dwarf
galaxies exhibit about 10 times higher radio-to-optical luminosity
than normal spirals and they have a flatter radio spectrum indicating
a weak synchrotron radiation (in turn possibly due to a weak magnetic
field). These findings have been largely confirmed by subsequent works
devoted to dIrrs (Klein \etal 1986) and larger samples (Klein \etal
1991).  This idea will also apply to larger starburst galaxies.

\acknowledgments We are grateful to E. Tolstoy, D. Bomans,
R.-J. Dettmar, G. Golla, and L. van Zee for discussions of the
observations.  D. Strickland \& I. Stevens kindly allowed us to quote
their results prior to publication.  Computations were performed at
the Rechenzentrum Garching of the Max-Planck-Gesellschaft.  Each author
thanks the other's institute for hospitality during work on this paper.


\newpage
\begin{center}
\Large Tables
\end{center}

\begin{table}[hp]
\begin{tabular}{lllllllll}
$M_g$ ($M_\odot$) & $10^6$ & $10^7$ & $10^8$ & $10^9$ & $10^6$ & $10^7$ & $10^8$ & $10^9$\\ 
\hline
$M_h$\tablenotemark{a} ($M_{\odot}$) & 6.8 (7) & 3.5 (8) & 1.8 (9) & 9.1 (10) & & & & \\
$v_c$ (km s$^{-1}$) & 19 & 34 & 58 & 99 & & & & \\
\hline
$c_{s,eff}$ ($10 \mbox{ km s}^{-1}$) & 1.0 & 1.0 & 1.0 & 1.0 & 0.6 & 0.6 & 0.6 & 0.6 \\ 
$H$ (kpc) & 0.64  & 0.59 & 0.55 & 0.51 & 0.23 & 0.21 & 0.20 & 0.18           \\ 
$n_0$\tablenotemark{a} (cm$^{-3})$ & 5.3 (-3) & 1.2 (-2) & 2.7 (-2) &
6.2 (-2) & 1.5 (-2) & 3.3 (-2) & 7.6 (-2) & 0.173 \\
$N_H$ ($10^{19}$~cm$^{-2}$)   & 1.0 & 2.2 & 4.6 & 9.8 & 1.0 & 2.2 & 4.6
& 9.8   \\
\end{tabular}
\caption{Galactic Parameters \label{pars}
\tablenotetext{a}{Factors of powers of 10 indicated in parentheses}
}
\end{table}



\begin{table}[hp]
\begin{tabular}{llll}
Visible Mass  & \multicolumn{3}{c}{Luminosity ($10^{38}$ erg
s$^{-1}$)} \\
($M_g/M_{\odot}$) & 0.1 & 1.0 & 10 \\
\hline
$10^6$ & 0.18     & 1.0      & 1.0       \\
$10^7$ & 3.5 (-3) & 8.4 (-3) & 4.8 (-2)  \\
$10^8$ & 1.1 (-4) & 3.4 (-4) & 1.3 (-3)  \\
$10^9$ & 0.0      & 7.6 (-6) & 1.9 (-5)  \\
\end{tabular}
\caption{Mass Ejection Efficiency, $\xi$ \label{mass}}
\end{table}

\begin{table}[hp]
\begin{tabular}{llll}
Visible Mass  & \multicolumn{3}{c}{Luminosity ($10^{38}$ erg
s$^{-1}$)} \\
($M_g/M_{\odot}$) & 0.1 & 1.0 & 10 \\
\hline
$10^6$ & 1.0  & 0.99 & 1.0  \\
$10^7$ & 1.0  & 1.0  & 1.0  \\
$10^8$ & 0.80 & 1.0  & 1.0  \\
$10^9$ & 0.0  & 0.69 & 0.97 \\
\end{tabular}
\caption{Metal Ejection Efficiency, $\xi_Z$ \label{metal}}
\end{table}

\clearpage

\begin{figure}
\caption{
Regions of the gas mass ($M_g$) -- 
mechanical luminosity of the starburst ($L$) plane in which 
blowout or blowaway can occur for a values of $c_{s,eff}=10$~km~
s$^{-1}$ and $h=0.65$.  The ratio $\epsilon$ of the galactic radius
$\varpi_*$ to the scale height $H$ is also shown.
\label{anal}   
} 
\end{figure}

\begin{figure}
\caption{Density distributions for models with the given initial
visible masses $M_g$ and luminosities $L$ at times of (a) 50, (b) 75,
(c) 100, and (d) 200 Myr, with the values of density given at the top
of each figure.  Note that energy input ends at 50 Myr. 
\label{den}    
}
\end{figure}

\begin{figure}
\caption{
Distribution of metal-enriched stellar outflow material and supernova
ejecta from the starburst energy source at a time of 200 Myr is shown in white,
superimposed on the same density distribution shown in Figure~2(d).
The material was followed with a tracer field, as described in the
text.  Note that, in most cases, no enriched gas remains in the disks
of the galaxies.
\label{col}
}
\end{figure}

\begin{figure}
\caption{ A direct computation of the temperatures on our grid at a
time of 100 Myr, assuming the gas is ionized with a number fraction of
0.1 of He.  These temperatures are qualitatively correct--hot gas in
our models should correspond to hot gas physically--but by no means
quantitatively correct, as we have neglected important microphysics
and injected an arbitrary amount of mass at the center to maintain the
superbubble interior density at roughly correct values.\label{temp}
  }
\end{figure}

\begin{figure}
\caption{ A qualitative image of the regions in our models likely to
be brighter in H$\alpha$ and other optical lines, at a time of 100
Myr.  We show the emissivity $I_{opt} \propto \rho^2 T^{-1/2}$ on
two-dimensional cuts through the galaxies corresponding to our
computational grid: these images have not been rotated and projected.
Note the prominence of filaments relatively close to the plane of the
disk even for extreme blowouts. \label{opt}
 }
\end{figure}

\begin{figure}
\caption{ A qualitative image of the regions in our models likely to
be brighter in X-ray emission at a time of 100 Myr.  We show an
emissivity $I_{opt} \propto \rho^2 T^{1/2}$ for $T > T_{min}$, as
discussed in the text.  Again, as in Figure~\ref{opt}, these are
two-dimensional cuts that have not been rotated and projected.  In
this image we neglect conductive evaporation off the denser filaments,
which may in fact dominate the observed X-ray emission. \label{xray}
 }
\end{figure}

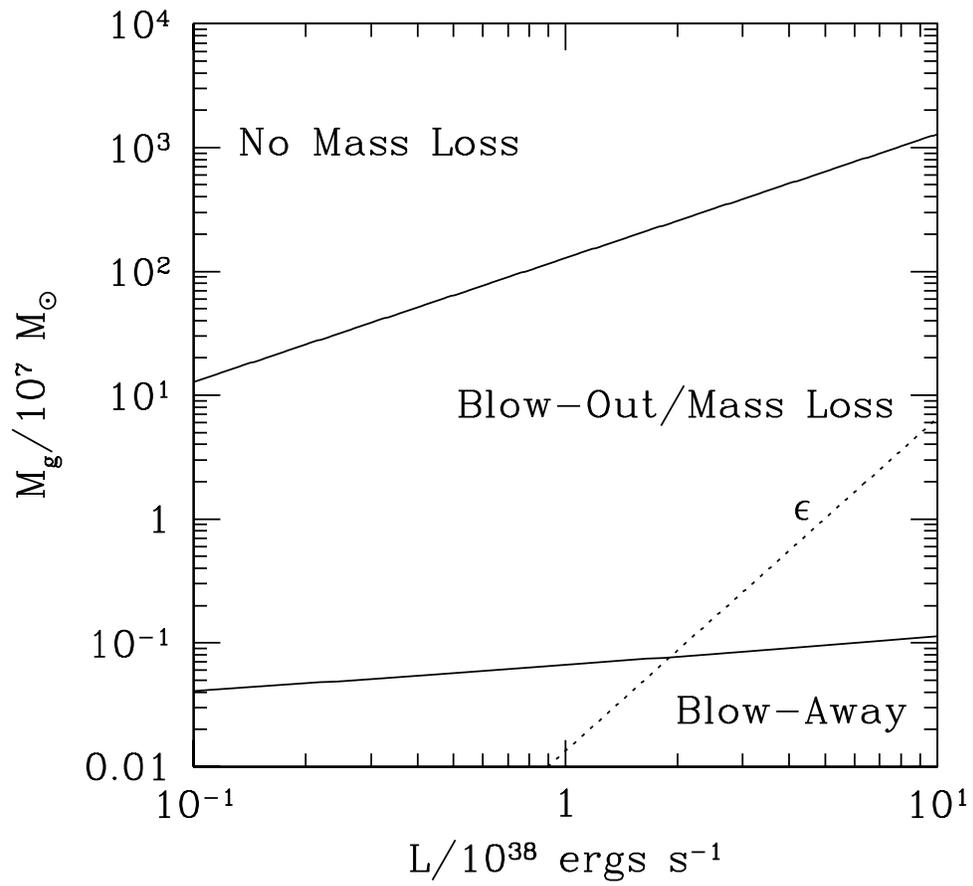
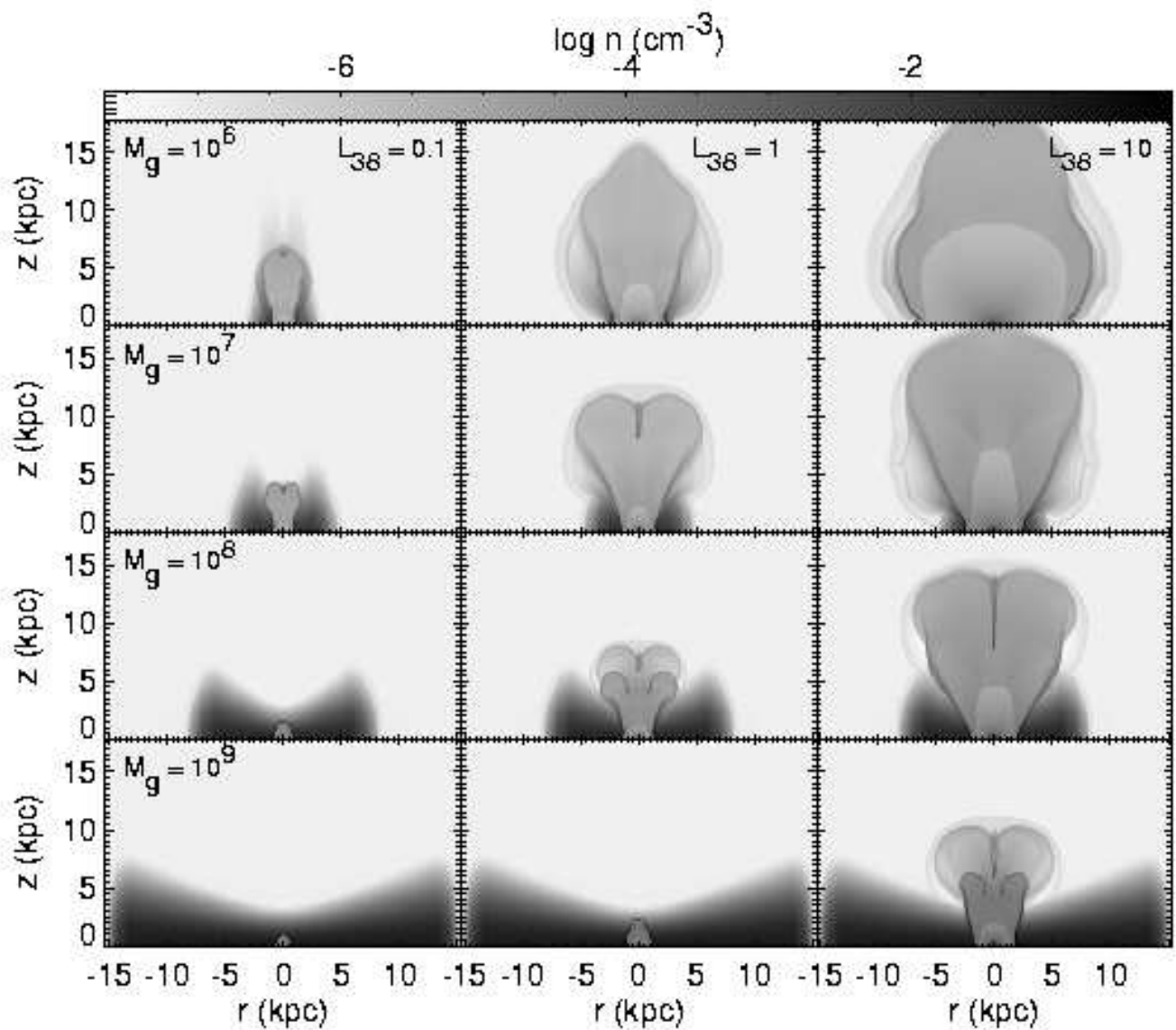
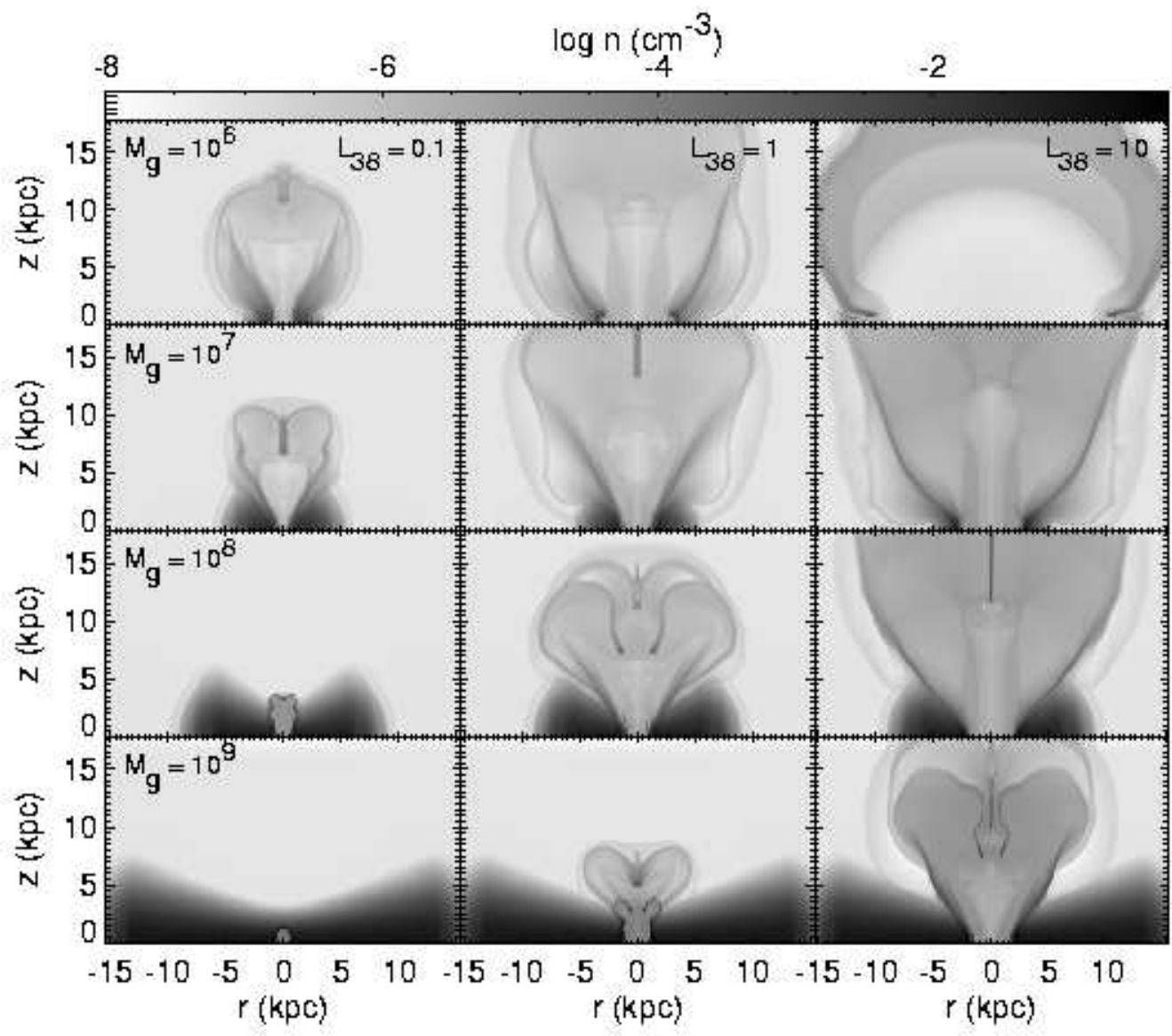
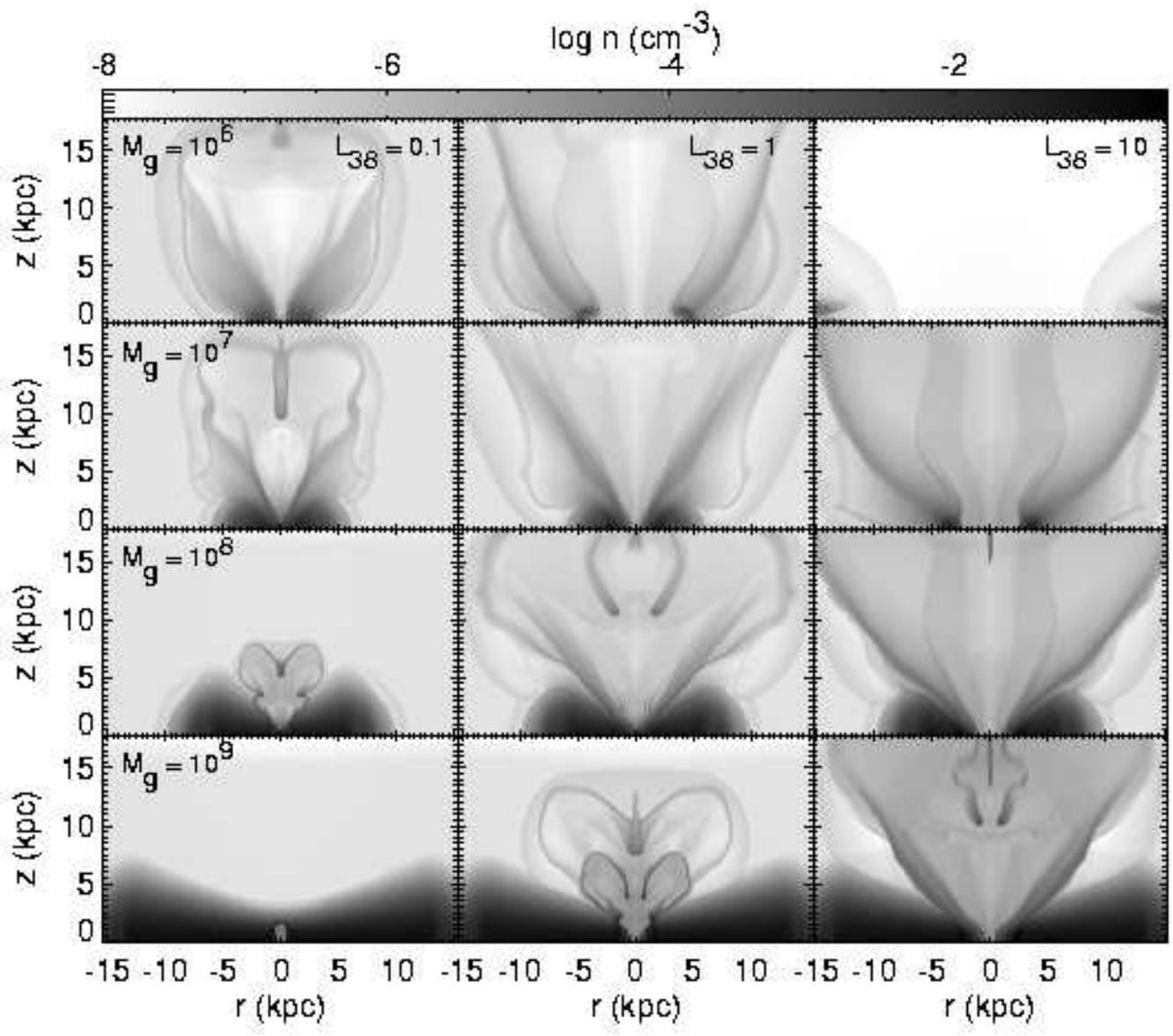
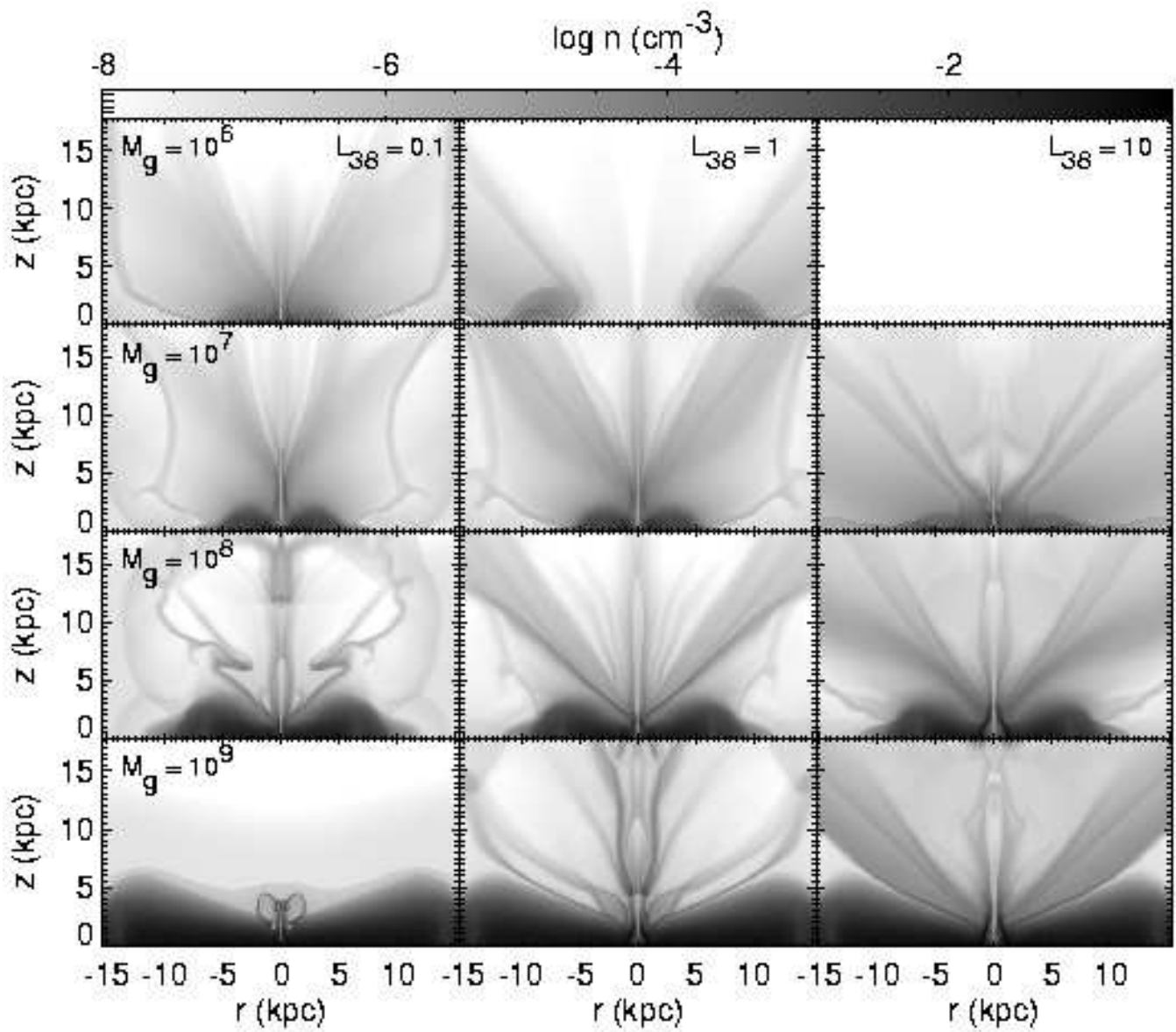
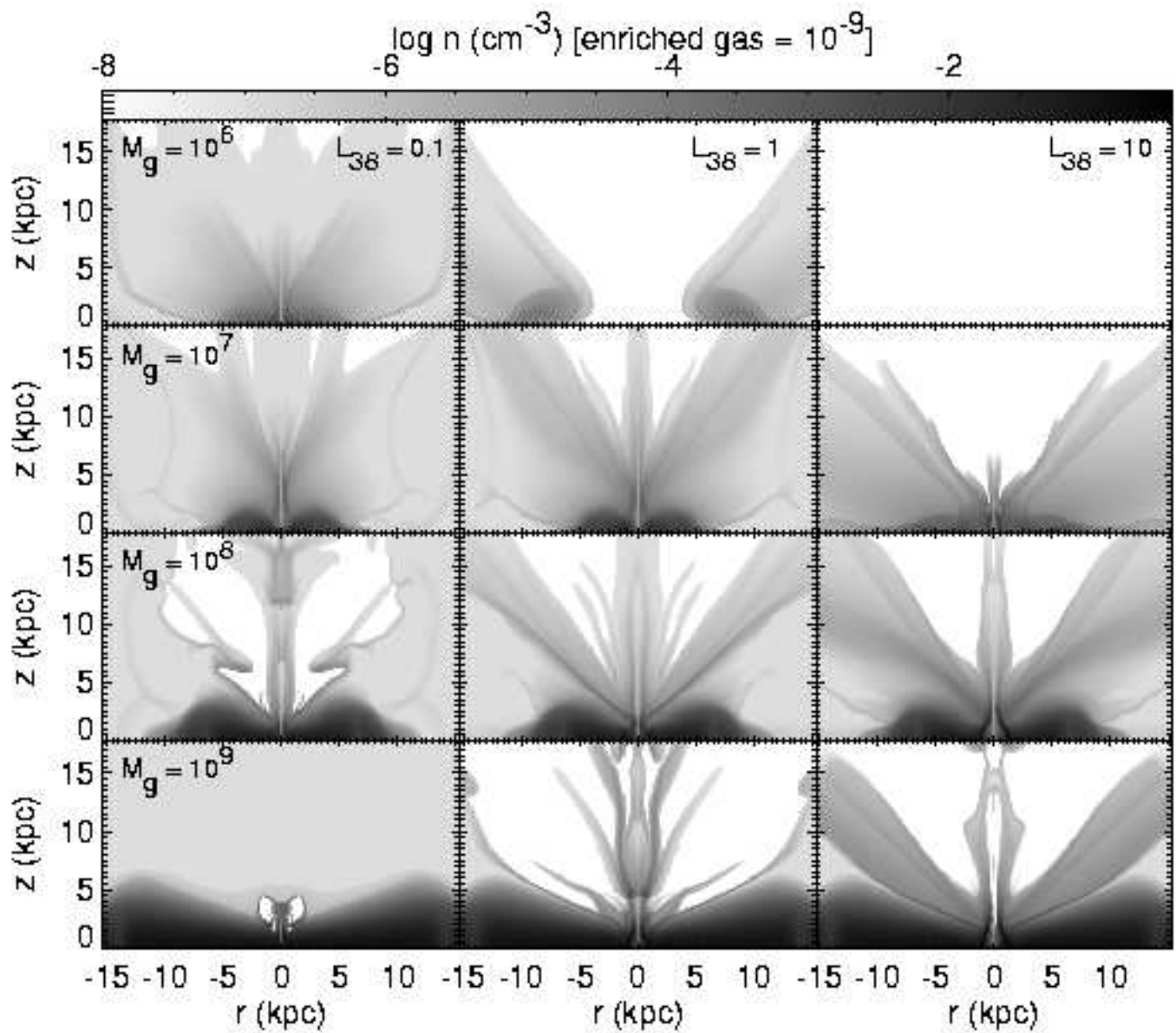
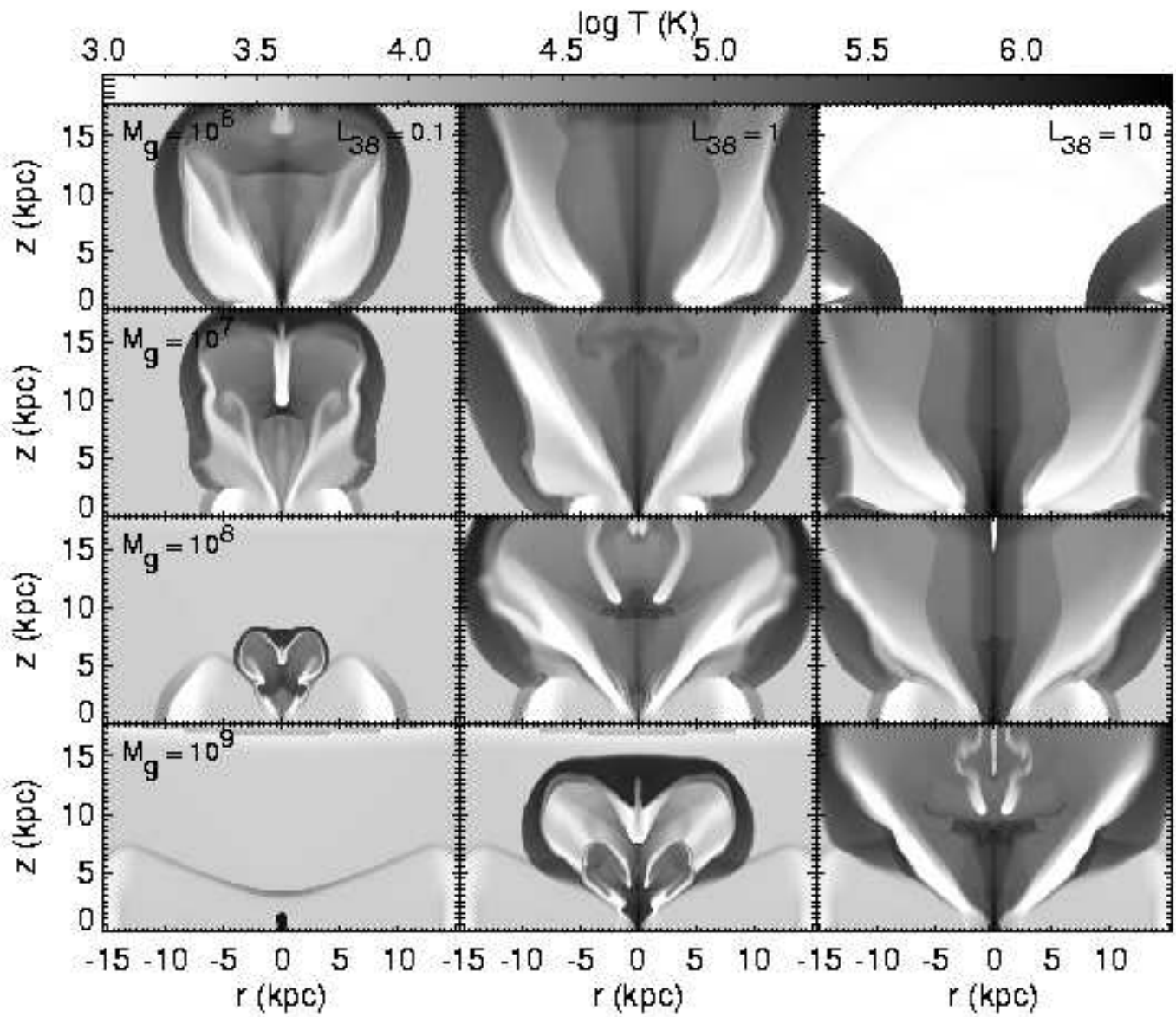
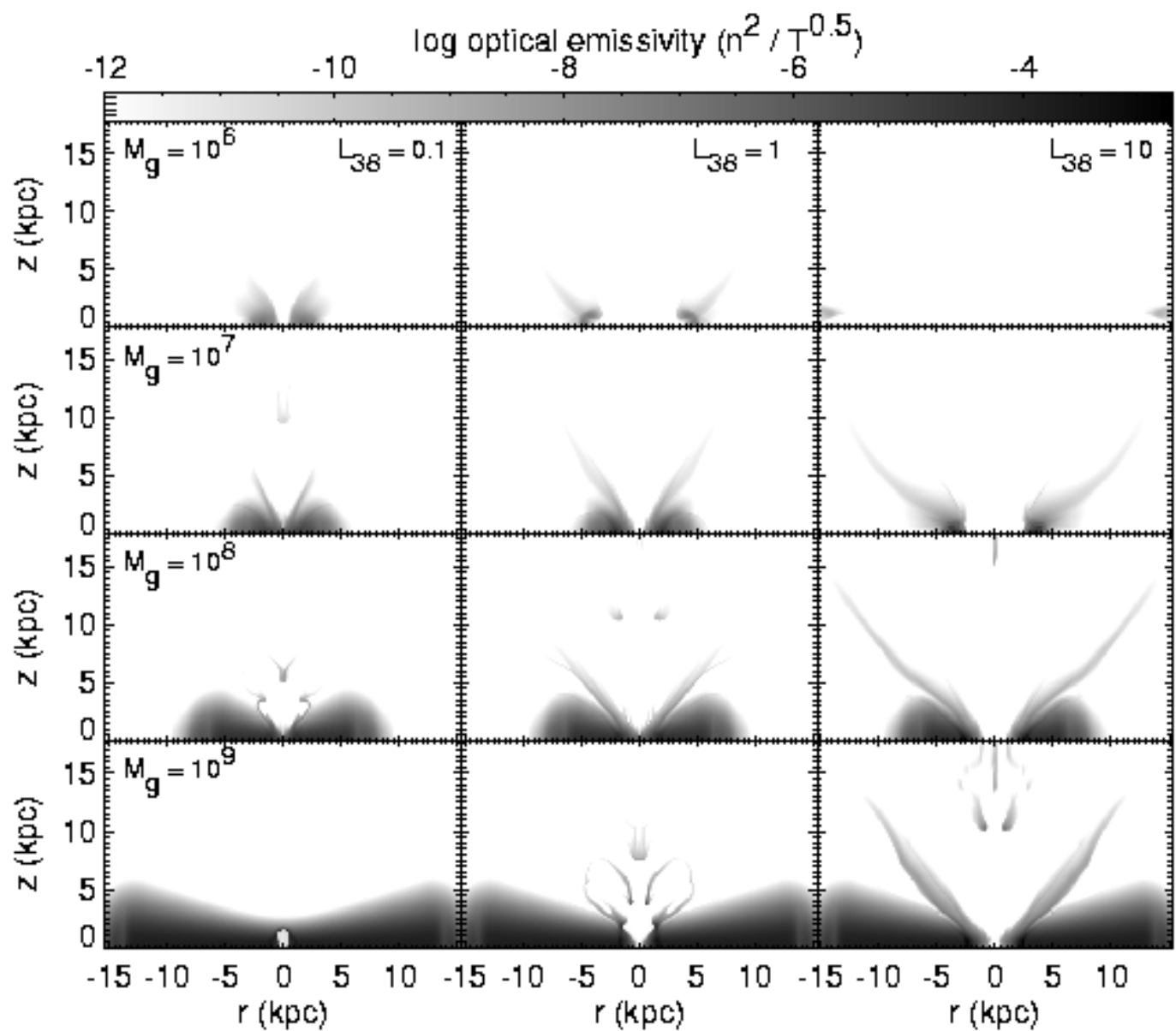
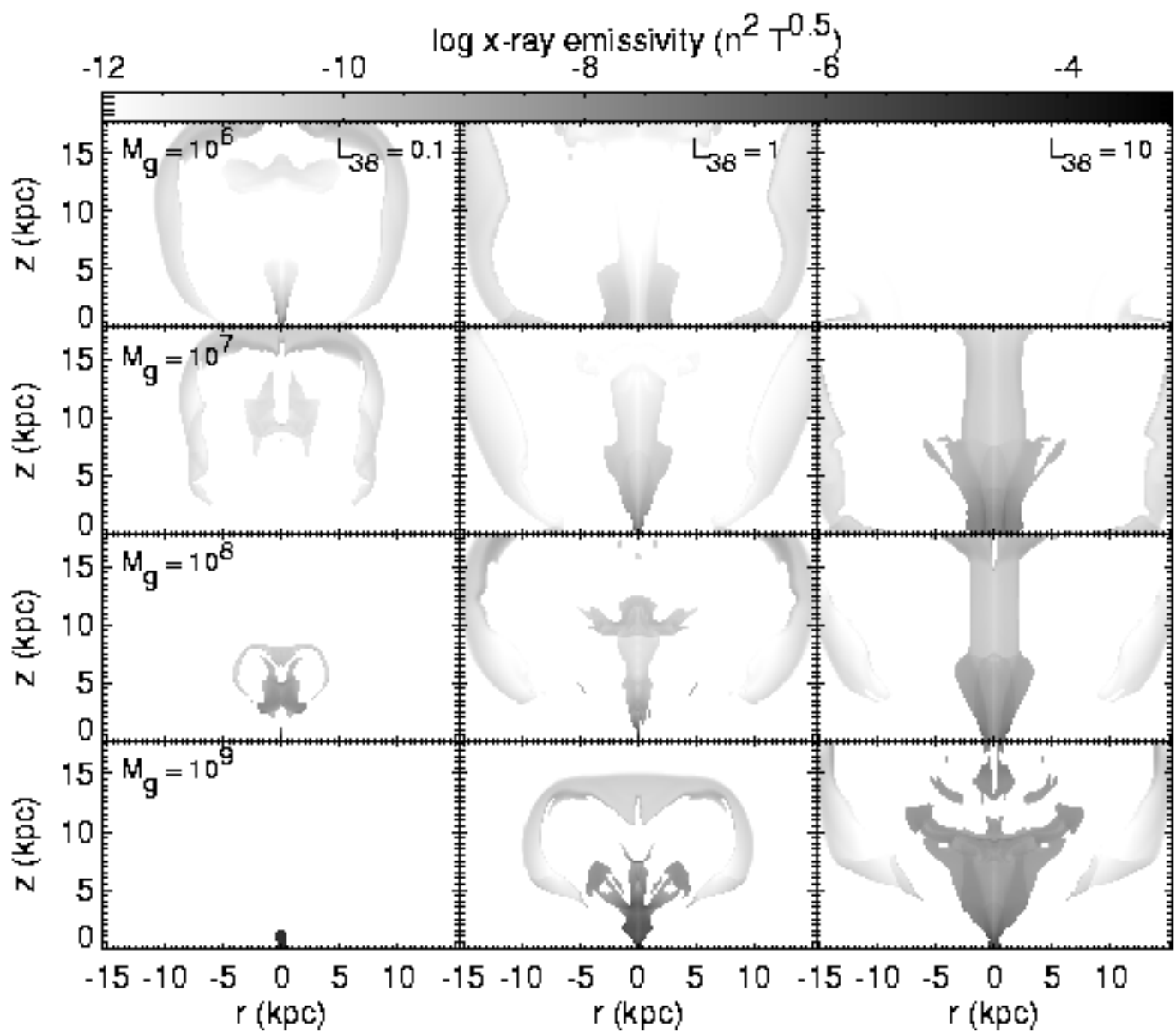
\end{document}